\documentclass[12pt]{article}
\usepackage{amsmath,amssymb}
\usepackage{latexsym,graphicx,epsfig,wrapfig}

\newcommand{\be}{\begin{equation}}
\newcommand{\ee}{\end{equation}}
\newcommand{\bes}{\begin{subequations}}
\newcommand{\ees}{\end{subequations}}
\newcommand{\bea}{\begin{eqnarray}}
\newcommand{\eea}{\end{eqnarray}}
\newcommand{\ba}{$$\begin{array}}

\newcommand{\ea}{\end{array}$$}

\newcommand{\bear}{\begin{equation}\begin{array}}

\newcommand{\fr}[2]{\dfrac{{ #1}}{{ #2}}}

\newcommand{\la}{\langle}
\newcommand{\ra}{\rangle}

\def\vep{{\varepsilon}}

\def\cl{\centerline}
\def\lb{\linebreak[4]}
\def\bu{$\bullet$}

\newsavebox{\fmbox}
\newenvironment{fmpage}[1]
{\begin{lrbox}{\fmbox}\begin{minipage}{#1}}
{\end{minipage}\end{lrbox}\fbox{\usebox{\fmbox}}}

{\end{list}}
\newcounter{enumct}

\allowdisplaybreaks 

\begin{document}
\title {\bf Discrete and continuous description of physical phenomena}

\author{I.~F.~Ginzburg\\
{\it Sobolev Institute of Mathematics, Novosibirsk, 630090, Russia;}\\  {\it Novosibirsk State University, Novosibirsk, 630090, Russia}}

\date{}
\maketitle

{\it \cl{To be published in}  \cl{\it Proc.
5 Symposium on Prospects
in the Physics of Discrete Symmetries,}
\cl{ "DISCRETE 2016",}
\cl{Warsaw  28.11 - 3.12, 2016.}

}

\begin{abstract}

The values of many phenomena in the Nature $z$ are determined in some discrete set of times $t_n$, separated by a small interval $\Delta t$  (which may also represent a coordinate, etc.). Let the $z$ value
in neighbour  point $t_{n+1}=t_n+\Delta  t$  be expressed by the
{\it evolution equation}
 as $z(t_{n+1})\equiv z(t_n+\Delta  t)=f(z(t_n))$. This equation gives {\it\underline{a discrete description}}  of phenomenon. Considering phenomena at $t\gg \Delta t$  this equation is transformed often into the differential equation allowing to determine  $z(t)$ -- {\underline{\it continuous description}}. It is  usually assumed that the continuous description
describes correctly the main features of a phenomenon at values $t\gg \Delta t$.

In this paper I show that  the real behavior of some physical
systems can differ strongly from that given by the continuous description.
The observation of such effects may lead to the desire to supplement the original evolutionary model by additional mechanisms, the origin of which require special explanation. We will show that such construction may not be necessary -- simple evolution model can describe different observable effects.

This text contains no new calculations. Most of the discussed facts are well known. New is the treatment of the results.

\end{abstract}


Let the  phenomenon  described by quantity $z_n\equiv z(t_n)$ be determined
in some discrete set of times $t_n$ with steps $\Delta t$ and  the $z(t_{n+1})\equiv z(t_n+\Delta  t)$  obey {\it the evolution equation}
\be
z(t_{n+1})\equiv z(t_n+\Delta  t)=f(z(t_n)).\label{basic1}
\ee
This  is {\underline{\it a discrete description}} of the phenomenon. This equation is transformed often into the differential equation for  $ z(t)\leftarrow z_n$, giving  {\underline{\it the continuous description}} \cite{Difeq}:
\be
\fr{z(t_{n+1})-z(t_n)}{\Delta t}\;\Rightarrow\;\fr{dz(t)}{dt}
=\fr{f(z(t_n))-z_n}{\Delta t}
\equiv f_1(z(t))\,.\label{cont1}
\ee
It is usually  assumed that the latter equation correctly describes the main features of the phenomenon at time $t\gg \Delta t$.

In this report I show that  in many cases the transition to the continuous description results in an incorrect description of reality, described by evolution equation \eqref{basic1}.  {\it Vise versa},
the {\it natural} interpretation of some observations can result in an incorrect description of elementary mechanisms.

\section{Model}

For definiteness, we consider the livestock of carps $z_n$ in a
pond without pikes ($n$ is the number of year). In this example
time spacing is limited from below, $\Delta t=1$ year (time between
spawning). The simplest model consider only natural growth,
$z_{n+1}=kz_n$ ($k$ is called {\it reproduction factor}). At $k>1$ this equation describe unlimited growth $z_n =k^n z_0$ .

In reality this growth is stopped by competition for food and
other effects ({\it pressure of population}). This pressure is
described by adding in our equation the term $-\ell z_n^2$.
As a result, our physical model takes the form of the evolution equation
 $ z_{n+1}=kz_n-\ell z_n^2$.
The change of variables $x_n=(\ell/k)z_n$ transforms this equation
in the more convenient form
 \be
x_{n+1}=f_1(x_n)\,,\qquad f_1(x)=kx(1-x)\,,\qquad x_n\ge
0\,.\label{basic}
 \ee

We are interested situation when the lifetime of population is not limited, the evolution starts from initial condition $x_0>0$ and  $k>1$.

Note that our problem has two equilibrium  solutions,
 \be
x_\infty=f(x_\infty)\;\Rightarrow\;\left\{
\begin{array}{ll c}
 x_{\infty,1}=0\;&(\mbox{\it stable at}\;&k<1)\,,\\
x_{\infty,2}=\dfrac{k-1}{k}\;&(\mbox{\it stable at}\;&k>1)
\end{array}\right.\,.\label{stac1}
 \ee

\section{Continuous description. Differential equation}

Ordinarily to describe phenomena at the scales larger than   $\Delta t$  the above
difference equation is conventionally reduced to the differential equation \cite{Difeq}. To
this end, the continuous  variable -- {\it time} $t$ is introduced instead of number of year  $n$. At
$t\gg 1$ our equation (\ref{basic}) is transformed to continuous
limit with the aid of standard sequence of relations \eqref{cont1} with $\Delta t=1$:
 \bear{c}
\dfrac{ x_{n+1}-x_n}{\Delta t}=x_n(k-1-kx_n)\;
 \Rightarrow\;\;
 \dfrac{dx}{dt}=x(k-1-kx)\,,\qquad x(0)=x_0\,.\end{array}\label{difeq}
 \ee

This differential equation is solved easily,
\be
x=\dfrac{(k-1)x_0}{kx_0+(k-1-kx_0)e^{-(k-1)\,t}}\;
.\label{soldifeq}
 \ee
This equation describes the evolution during infinite time  for an {\bf arbitrary value of $\pmb k$} and {\bf on arbitrary initial value $\pmb {x(0)>0}$}. At $t\to\infty$ and $k>1$ the quantity $x$ tends monotonically to the limiting value    (\ref{stac1}) $x_{\infty,2}=(k-1)/k$, determined for the incident evolution equation.

\section{Real evolution. Discrete description}

Below we reproduce some known results of the study   of evolution model \eqref{basic} (see e.g. \cite{Feig}). Our task is to present some badly known interpretations.

\bu \ In contrast to the  continuous description
the equation \eqref{basic} describes evolution of system during infinite time
 only at  $0<x_n<1$, in particular at
 \be
 0<x_0<1\,. \label{region}
 \ee
Indeed, at $x_n>1$ the eq.~(\ref{basic})
gives senseless $x_{n+1}<0$,  the process is stopped.

\bu \ The details of picture are different at different values of the reproduction factor $k$.

\subsection{{The case $1<k<3$.} }

We
consider the stability of the asymptotical solution (\ref{stac1}) at large time
$n$. To this end, we substitute in eq.~(\ref{basic}) the
expression $x_n=x_\infty +\delta_n\equiv (k-1)/k+\delta_n$ and,
considering $\delta_n$ to be small at large $n$, linearize this
equation (neglecting terms $\propto\delta_n^2$)). It results in the
equation for $x_\infty$ and the recurrence relation for $\delta_n$,
 \be
x_\infty =kx_\infty(1-x_\infty)\,,\qquad
\delta_{n+1}=(2-k)\delta_n\,.\label{kless3}
 \ee

 The equation for $x_\infty$ has
two solutions (\ref{stac1}). At $k<1$ the trivial solution
$x_\infty\equiv x_{\infty,1}=0$ is realized (by the population going extinct). At $k>1$
nontrivial solution $x_\infty\equiv x_{\infty,2} =(k-1)/k$ is realized. The relation
for  $\delta_n$ shows that at $|2-k|<1$ (i.e. at $1<k<3$) the
quantity $\delta_n$ tends to 0 at $n\to\infty$, and the solution
converges to $x_\infty\equiv x_{\infty,2}$.

$\lozenge$ At $1<k<2$
the quantity $x_n$ tends to grow monotonically to $x_{\infty}$  for larger $n$
just as the solution of differential
equation \eqref{soldifeq}. Continuous description works for initial values \eqref{region}.

$\lozenge$ At $2<k<3$ the quantity $x_n$ also tends to $x_\infty$
but in contrast with the solution of the differential
equation, this increase is not monotonic but (damped) oscillating.
The continuous description describes the  main features of evolution (limiting value) but not in important details.

\subsection{{The case $3<k<3.45$. Period doubling.}}

According to eq.~(\ref{kless3}) at
$k>3$ we have $|\delta_{n+1}|>|\delta_n|$. Therefore, the
asymptotic solution (\ref{stac1}) becomes unstable. The point $k=3$
is a branching point. What next?

Let us check "the simplest" hypothesis that  at $n\to\infty$
the number of carps is reproduced through two years (not yearly),
with different populations in even and odd years. In other words,
we suggest that $k=3$ is the point  of the doubling of the
period.

The iteration of eq.~(\ref{basic}) gives the equation
 \be
x_{n+2} =f_2(x_n),\quad f_2(y)=f_1(f_1(y))\equiv
k^2y(1-y)[1-ky(1-y)]\,.\label{2discreq}
 \ee
The possible stationary states are given by equation of 4-th order
$x_\infty-f_2(x_\infty)=0$. Two solutions of this equation are
also known solutions (\ref{stac1}) of equation $x_\infty-f_1(x_\infty)=0$. To find other solutions we consider equation
$[x_\infty-f_2(x_\infty)]/[x_\infty-f_1(x_\infty)]=0$ which has
the form $k^2y^2 - k(k + 1)y + k + 1 = 0$. Its solution exists
only at $k>3$:
  \be
y=x_{\infty,3,4} \equiv\fr{k+1\pm\sqrt{(k+1)(k-3)}}{2k}\Rightarrow \left\{\begin{array}{c}
f_1(x_{\infty,3})=x_{\infty,4}\,,\\[2mm]
f_1(x_{\infty,4})=x_{\infty,3}\,.
\end{array}\right.
 \label{2stac}
 \ee
It is easy  to check that
$$
x_{\infty,3}>x_{\infty,2}=(k-1)/k>x_{\infty,4}\, .
$$

Therefore, at large $n$ the quantity $x_n$ oscillates between the values
$x_{\infty,3}$ and $x_{\infty,4}$ with a period of 2 years. This phenomenon is called the {\it period doubling}.

Now we find the range of validity of this. For this goal
we repeat  above calculations  with substitution $x_n=x_{(\infty,3,4)} +\delta_n$ in Eq.~\eqref{2discreq}. The obtained
linearized recurrence relation for $\delta_n$ has the form
 \be
 \delta_{n+2}=(4+2k-k^2)\,\delta_n\,.\label{2delta}
 \ee
Just as above we find that the solutions $x_{\infty,3,4}$ (\ref{2stac})  are
stable only at $|4+2k-k^2|<1$, i.~e. at $3<k<1+\sqrt{6}\approx
3.45$ only.

\subsection{{The case $3.45<k<4$.}}

\begin{wrapfigure}[12]{r}[0pt]{10cm}\vspace{-7mm}
\includegraphics[width=9cm]{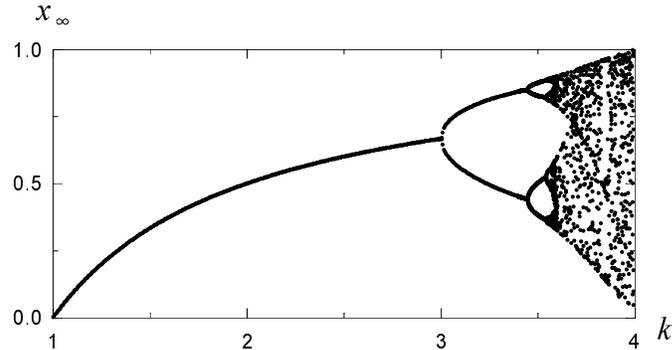}
\caption{\it Limiting values  $x_\infty$ at different $k$. }
   \label{figchaos}
    \end{wrapfigure}
With the growth of $k$ the analytical study of the problem becomes complex. The Fig.~\ref{figchaos}
(obtained with "Mathematica" package) represents
complete picture  of limiting values $x_\infty$ at different $k$.
 Here in the abscises axis the
reproduction factor $k$ is shown and at the ordinate axis --
values $x_\infty$ (unfortunately, details at $k>3.5$ are badly
seen).

At $1<k<3$ the single limiting value is given by eq.~(\ref{stac1}). At
$3<k<3.45$ the population oscillates between two values (the period doubling).
At $3.45<k<3.54$ we have a subsequent doubling of the period;
the population varies periodically between four values. With the
growth of $k$  the number of limiting values grows (new period doublings  replace each other).  At
$k>3.569$ zones of chaotic behavior appears. At
$k\to 4$ these zones are merged in the entire interval $(0,\,1)$.

\subsection{{The case $k=4$. Chaotic behaviour.}}

At $k=4$ evolution equation \eqref{basic} can be solved exactly.
We seek a solution in the form
 \be
x_n=\dfrac{1-\cos 2\pi\alpha_n}{2}\,.\label{ans}
 \ee
Now eq.~(\ref{basic}) gives a simple sequence of identities:
 $$
x_{n+1}=4\,\dfrac{1+\cos 2\pi\alpha_n}{2}\,\dfrac{1-\cos
2\pi\alpha_n}{2}=\sin^2(2\pi\alpha_n)=\dfrac{1-\cos
4\pi\alpha_n}{2}\,.
 $$
Comparing with (\ref{ans}), we find
$
\alpha_{n+1}
=2\alpha_n\;\Rightarrow\;\;\alpha_n=2^n\;\alpha_0$.
 The adding  of any integer number to $\alpha_n$ does not change $x_n$.
Therefore, in this solution  only $\{\alpha_n\}$
-- fractional part of $\alpha_n$ -- makes sense, and
\be
\alpha_n=\left\{2^n\;\alpha_0\right\}\,.\label{sol4}
\ee

Now we discuss the meaning of  this solution from different perspectives.

$\lozenge$ Let us consider two initial values $x_0(\alpha_0)$ and $x'_0(\alpha'_0)$ with $|\alpha_0-\alpha'_0|=\vep\ll 1$  and $2^{-r}>\vep>2^{-(r+1)}$. The
"respectable"\ solution should be stable, it means that it is
naturally to hope that the solutions with initial values $x_0$
and $x'_0$ are close to each other even after long time. On the
contrary, in our case at $n>r$ (after $r$ years) the difference
between $x_n$ and $x'_n$  becomes unpredictably large.

$\lozenge$ The variable $z_n$ is the number of carps averaged over
some period. It varies weakly due to natural and accidental
deathes. Therefore, the initial value $z_0$ (or $x_0$ or
$\alpha_0$) is determined with some uncertainty $\vep$ with
$2^{-r}>\vep>2^{-(r+1)}$. Let us present this initial state in the
binary-decimal form like $\alpha_0=0.1101000100...$. Here the $r$-th
term is known while the\lb $r+1$-th term is unknown. Solution
(\ref{sol4}) means that in each next year $\alpha_{n+1}$ is
obtained from $\alpha_n$ by shift of point from one digit to the
right with elimination of signs before the point. Therefore, the
population of carps is predicted via an initial state (by means of
solution (\ref{sol4})) during first $r$ years (with downward
accuracy) but the population in $r+1$-th year is unpredictable.

$\lozenge$ The typical initial value $\alpha_0$ is irrational. By
definition of irrational, the numbers -- consecutive values
$\{\alpha_n\}$ are non correlated. In other words, for almost all
randomly chosen $x_0$ (or $\alpha_0$) the range of produced values
$\{\alpha_n\}$ is distributed uniformly in the segment $[0,\,1]$.

$\blacklozenge$ Therefore, at $k=4$ the limiting value
$x_{n\to\infty}$ does not exist. The values $x_n$ vary with time
unpredictably.

Note
that even the averaged  value $\la x_n\ra=1/2$ differs
from the value 3/4 given by eq.~ (\ref{difeq}), see \cite{distrib}.

\subsection{{The case $k>4$.}}

 At $0<x_n<1$ the maximal value of the quantity $x_n(1-x_n)$
is 1/4. Therefore, to prevent the restraining of process at some
stage, one should be $k<4$. We expected that the mean time of life of population at $k>4$ is small, but at small values of $k-4$  it is not very small. The numerical experiment shows infinitesimal time of life even for $k-4=0.01$.

(Certainly, for some rational values $x(0)$ time of life can be large, but the rounding errors kill this opportunity).

\section{Discussion}

In two tables below we compare  some features of discrete and continuous description of our system in the regions of parameters, allowing evolution during infinite time. Most of mentioned facts  were discussed earlier \cite{Feig} but with another accents. We discuss lessons from this picture for two fields of studies.
\\

\cl{\begin{fmpage}{0.76\textwidth}
$$
\begin{array}{|l||c|c|}\hline
\mbox{description:}&\mbox{\it discrete }&\mbox{\it continious}\\ \hline \hline
\mbox{reprfoduction factor }&\mbox{restricted, $0<k\le 4$}&\mbox{arbitrary, $k>0$}\\
\hline
\mbox{initial values}&\mbox{restricted},\;\; 0< x(0)<1 &\mbox{arbitrary, $x(0)>0$}\\ \hline
\end{array}
$$
\cl{\bf Allowed ranges of reproduction factor and initial value.}
\end{fmpage}\\[5mm]}

\cl{\begin{fmpage}{0.98\textwidth}
$$\begin{array}{|l||c|c|}\hline
 k\downarrow\diagdown\;{\mbox{description:}\rightarrow}&\mbox{\it discrete}&\mbox{\it continuous }\\\hline\hline
  k<2&\mbox{monotonic approaching to}\; x_{\infty,2}&\mbox{monotonic}\\\hline
2<k<3&\mbox{ non-monotonic approaching to}\; x_{\infty,2}&\mbox{monotonic}\\ \hline
3<k<3.45& \mbox{oscillations between 2 values}&\mbox{monotonic}\\\hline
3.45<k<3.569& \mbox{oscillations between $2^n$ values}
 &\mbox{monotonic}
\\\hline
3.569<k<4& \mbox{unpredictable, chaotic behaviour appear}&\mbox{regular evolution}\\ \hline
k=4& \mbox{unpredictable, chaotic behaviour}&\mbox{regular evolution}\\ \hline
k>4&\mbox{Population is dying out}&\mbox{regular evolution}\\\hline
\end{array}
$$
\cl{\bf Behaviour at large time, $\pmb{t\equiv n\gg 1}$.}
\end{fmpage}
}
.\\


\bu \ \underline{\bf Lessons for using of  continuous approach for description of  real system.}

{\it The standard approach in the study of physical phenomena is to change discrete evolution equation for continuous one and subsequent using  of differential equation for the description of long-time evolution of system. Our discussion demonstrates that this approach is valid only for limited range of parameters.}

$\lozenge$ The approximation of discrete variation of some physical
quantity $x_{n+1}=f(x_n)$ by continuous one is justified only at
relatively low temp of variation of this quantity at one step
$k=\left[(x_{n+1}-x_n)/x_n\right]$ at $x_n\to 0$ in the basic difference equation. With growth
of this temp $k$ the behavior of such system can be changed
strong.

$\lozenge$ There are the threshold values $k_{0i}$ of reproduction factor $k$. The
differential equation describes well the behavior of physical
quantity $x$ at $k<k_{01}$ while at $k>k_{01}$ such description become
incorrect. The quantities $k_{0i}$ are different for different physical
properties. In our example\\ $\triangledown$ At $k<k_{01}(=2)$ continuous approach describes real picture well.\\ $\triangledown$ At $k_{01}<k<k_{02}(=3)$ continuous approach gives good description of main picture with incorrect description of important details (it skips non-monotonic variation of $x(t)$).\\
$\triangledown$  At $k>k_{02}(=3)$ the continuous approach absolutely
inapplicable for description of system during long time.

$\lozenge$ For the validity of continuous approach in the multi-dimensional problems the eigenvalues of reproduction factor matrix $\{k\}$ should be small enough (this limitation can be broken even at small values of separate elements of  matrix $\{k\}$ due to large number of its component.)

\bu \ \underline{\bf Lessons for construction of model for physical phenomenon.}

{\it Important problem is the construction of model for physical phenomenon, based on the observations and the preliminary understanding of the process.} Typical procedure is construction of evolution equation, its description in the continuous approach, and obtaining of coefficients from observation. If observed behaviour differs from that predicted in this approach not very strong, we conclude that our model is correct in general, but some additional  mechanisms should be taken into account for description of details. If we observe new phenomena, which don't appear in our continuous approach, we conclude that our incident understanding of problem was incorrect, and new mechanisms are responsible for this phenomenon.

Our analysis shows that such approach can be often incorrect. Simple model describes phenomenon, and only continuous approach can appear wrong.\\

{\bf Examples}.

$\lozenge$ {\it Let population dies out in final time.} In the traditional approach
it would be natural to seek an explanation in the existence of  some new force which kills population. We understand that in the discrete approach it will be due to inappropriate initial conditions or parameters (e.~g. $x_n>1$ or $k>4$) with the same simple evolution equation. ({\it In economics it corresponds to the development of the crisis at abnormally fast elementary growth.})

$\lozenge$  {\it Let we observe periodic behavior.} In the traditional approach
it would be natural to seek an explanation in the existence of   {\it additional} mechanism providing periodicity (periodic external force or resilient restoring force). Our example for $3<k<3.569$ shows that even complex periodic behaviour can be explained without these additional mechanisms.

$\lozenge$ {\it Let we observe chaotic  behavior.} In the traditional approach
it would be natural to seek an explanation in the existence of   {\it additional} mechanisms, like random external force (open system). Our example for $4>k>3.569$ shows that even complex chaotic behaviour can be explained without these additional mechanisms.

\section*{Acknowledgments}
I am thankful to I.P. Ivanov, G.L. Kotkin, M. Krawczyk and V.G. Serbo   for discussions. This work was supported in part by grants RFBR 15-02-05868 and HARMONIA project under contract
  UMO-2015/18/M/ST2/00518 (2016-2019).

\smallskip


\begin{thebibliography}{100}
\bibitem{distrib} At $k=4$ the probability $p(y)dy$
to find  $x_n$  within the interval $(y,\,y+dy)$ is
calculated with the aid of eq.~(\ref{ans}). For almost all initial
values $x_0$ we have\lb $p(y)=(1/\pi)\left[ y(1-y)\right]^{-1/2}$ with
 mean value $\la x_n\ra=1/2$ and fluctuation $\la
\Delta x_n\ra= \la x_n\ra$.
\end{thebibliography}
\end{document}